\documentclass[amsmath,amssymb,aps,prc,twocolumn,superscriptaddress,showpacs,preprintnumbers]{revtex4}
\usepackage{graphicx,color}
\usepackage{multirow}
\usepackage{times}
\usepackage{bm}
\usepackage{epstopdf}
\usepackage{enumerate}
\usepackage[normalem]{ulem}  % \sout{old text} for strikeout

\newcommand{\comment}[1]{}
\renewcommand\sout{\bgroup \color{red} \ULdepth=-.5ex \ULset}
\newcommand{\srtNN}{\sqrt{s_{{\scriptscriptstyle NN}}}}

\begin{document}
%\preprint{YITP-17-80}
\preprint{J-PARC-TH-0131}

\title{
Dynamically integrated transport approach
for heavy-ion collisions at high baryon density
}

\author{Yukinao Akamatsu}
\affiliation{Department of Physics, Osaka University, Toyonaka, Osaka
560-0043, Japan}
\author{Masayuki Asakawa}
\affiliation{Department of Physics, Osaka University, Toyonaka, Osaka
560-0043, Japan}

\author{Tetsufumi Hirano}
\affiliation{Department of Physics, Sophia University, Tokyo 102-8554, Japan}

\author{Masakiyo Kitazawa}
\affiliation{Department of Physics, Osaka University, Toyonaka, Osaka
560-0043, Japan}
\affiliation{J-PARC Branch, KEK Theory Center,
Institute of Particle and Nuclear Studies,
KEK, 203-1, Shirakata, Tokai, Ibaraki, 319-1106, Japan}

\author{Kenji Morita}
\affiliation{Institute of Theoretical Physics, University of Wroclaw,
50204 Wroc{\l}aw, Poland}
\affiliation{iTHES Research Group, RIKEN, Saitama 351-0198, Japan}

\author{Koichi Murase}
\affiliation{Department of Physics, The University of Tokyo,
7-3-1 Hongo, Bunkyo-ku, Tokyo 113-0033, Japan}
\affiliation{Department of Physics, Sophia University, Tokyo 102-8554, Japan}

\author{Yasushi Nara}
\affiliation{Akita International University, Yuwa, Akita-city 010-1292, Japan}
%\affiliation{Frankfurt Institute for Advanced Studies, 
%D-60438 Frankfurt am Main, Germany}

\author{Chiho Nonaka}
\affiliation{Department of Physics, Nagoya University, Nagoya 464-8602, Japan}
\affiliation{Kobayashi Maskawa Institute, Nagoya University, Nagoya 464-8602,
Japan}

\author{Akira Ohnishi}
\affiliation{Yukawa Institute for Theoretical Physics, Kyoto University,
Kyoto 606-8502, Japan}
%%%%%

\date{\today}

\begin{abstract} 
We develop a new dynamical model for high-energy heavy-ion collisions 
in the beam energy region of the highest net-baryon densities
on the basis of nonequilibrium microscopic transport model JAM
and macroscopic (3+1)-dimensional hydrodynamics
by utilizing a dynamical initialization method.
In this model, 
dynamical fluidization of a system is controlled by
the source terms of the hydrodynamic fields.
In addition, time dependent core-corona separation of hot regions
is implemented.
We show that our new model describes multiplicities and mean transverse mass
in heavy-ion collisions within a beam energy region of
$3<\sqrt{s_{NN}}<30$ GeV.
Good agreement of the beam energy dependence of
the $K^+/\pi^+$ ratio is obtained,
which is explained by the fact that
a part of the system is not thermalized in our core-corona approach.
\end{abstract}

\maketitle

\section{Introduction}

The study of the structure of the QCD phase diagram is one of the most
important subjects in high-energy heavy-ion physics.
Ongoing experiments such as the Relativistic Heavy-Ion Collider (RHIC) Beam
Energy Scan (BES) program~\cite{Adamczyk:2017iwn,Aggarwal:2010cw,Kumar:2013cqa}
and the NA61/SHINE experiment at the Super Proton Synchrotron
(SPS)~\cite{NA61SHINE}
enable us to explore the high baryon density domain
by creating compressed baryonic matter (CBM)
in laboratory experiments~\cite{cbmbook}.
Future experiments currently planned such as,
BES II of STAR at RHIC~\cite{BESII},
the CBM experiment at FAIR~\cite{FAIR},
MPD at NICA,  JINR~\cite{NICA},
and a heavy-ion program at J-PARC (J-PARC-HI)~\cite{HSakoNPA2016}
will offer opportunities at the most favorable beam energies
to explore the highest baryon density matter.
These studies on the high-baryon density matter may also have implications
for understanding neutron stars and their mergers
through astronomical observations~\cite{Baym:2017whm}.
For example, the binary neutron-star mergers~\cite{TheLIGOScientific:2017qsa}
are expected to discriminate dense baryonic matter
equation of state (EoS) at
the densities five to ten times higher than the normal  nuclear matter
density
$\rho \sim (5-10) \rho_0$
($\varepsilon=(0.75-1.5)~\mathrm{GeV}/\mathrm{fm}^3$)
via the gravitational wave spectrum~\cite{Sekiguchi:2011zd}.
In heavy-ion collisions,
the search for a first-order phase transition and
the QCD critical point predicted by some theoretical models%
~\cite{DHRischke2004}
is one of the most exciting topics.
 
To extract information from experiments, we need dynamical models
for understanding the collision dynamics of heavy-ion collisions. 
At RHIC and the LHC, the hydrodynamical description of heavy-ion collisions
has been successful in explaining a vast body of data, in which
hydrodynamical evolution starts at a fixed proper time of about
$\tau \sim 1$ fm/$c$ with initial conditions provided by
some other theoretical models~\cite{Heinz:2013th,Gale:2013da,
Huovinen:2013wma,Hirano:2012kj,Jeon:2015dfa,Jaiswal:2016hex,Romatschke:2017ejr}.
Systematic analyses of RHIC and LHC data have been done based on Bayesian
statistics using state-of-the-art hybrid simulation codes
to extract the quark gluon plasma (QGP) properties
~\cite{Pratt:2015zsa,Bernhard:2016tnd}.
However, this picture breaks down at lower beam energies $\sqrt{s_{NN}}<30$ GeV
for the description of heavy-ion collisions
since the passing time of two nuclei exceeds 1 fm/$c$ and secondary
interactions become important before two nuclei pass through each other.
Thus one needs a nonequilibrium transport model to follow
dynamics before hydrodynamical evolution starts.
We note, however, that as an alternative approach to heavy-ion collisions
at baryon stopping region, three fluid dynamics (3FD) has
been used extensively to analyze the collision dynamics
\cite{Brachmann:1997bq,Ivanov:2005yw,Ivanov:2013yla,Batyuk:2016qmb,Ivanov:2018vpw}.

The UrQMD hybrid model has been developed for 
simulations of heavy-ion collisions at high baryon density%
~\cite{Steinheimer:2007iy,Petersen:2008dd}.
In this approach,
the initial nonequilibrium dynamics of the collision is treated by the UrQMD model,
and one assumes that the system thermalizes just after two nuclei
pass through each other, then dynamics of the system
is followed by hydrodynamics. Finally, after the system
becomes dilute, one switches back to UrQMD to follow the time evolution
of a dilute hadron gas.
It was pointed out that the separation of the high density (core)
and the peripheral (corona) part is significant
at the top SPS and RHIC energies for the description of
centrality dependence of the nuclear collisions~\cite{Werner:2007bf}.
As core-corona separation should be also significant at lower beam
energies, it has been implemented into a UrQMD hybrid model,
which improves the description of the experimental data%
~\cite{Steinheimer:2011mp}.
This approach has been extended by incorporating viscous hydrodynamics
\cite{Karpenko:2015xea,Auvinen:2017fjw}.

Recently, a dynamical initialization approach has been proposed
on the basis of the hydrodynamics with source terms
which enable a unified description of
energy loss of jets by the bulk hydrodynamic environment
at RHIC and LHC energies~\cite{Okai:2017ofp}.
A similar idea was applied to develop a new dynamical
initialization approach based on string degrees of freedom
for the description of heavy-ion collisions at RHIC-BES
energies~\cite{Shen:2017bsr}.
In this approach, instead of assuming a single thermalization
time, hydrodynamical evolution starts at different times locally,
so one can simulate a collision of finite extensions
of colliding nuclei.

In this paper, we present a new dynamically integrated transport model
by combining the JAM transport model and hydrodynamics.
We utilize the same idea as Ref.~\cite{Okai:2017ofp,Shen:2017bsr}
for the dynamical initialization of the fluids. 
At the same time,
our approach takes into account the core-corona separation picture
both in space and time;
hydrodynamical evolution starts at different spacetime
points where energy density is sufficiently high.
Another distinct feature of our approach
from the previous models~\cite{Okai:2017ofp,Shen:2017bsr}
is that spacetime evolution of nonequilibrium part of the system is 
simultaneously solved by the microscopic transport model
together with hydrodynamical evolution.

Most of the hadronic transport models lack multiparticle interactions
in the dense phase. Hybrid approaches can overcome this type of defect.
An immediate consequence of the improvement is the enhancement of
the strange particle yields relative to the predictions
by a standard hadronic transport approach,
as pointed out in Ref.~\cite{Petersen:2008dd}.

Statistical model predicts the nontrivial structure~\cite{Gazdzicki:1998vd}
called ``horn'' in the excitation function of the $K^+/\pi^+$ ratio,
which was observed in the experiments~\cite{Afanasiev:2002mx,Alt:2007aa}.
The existence of this sharp structure has been explained by
several statistical models %
~\cite{Andronic:2005yp,Andronic:2008gu,Satarov:2009zx,Becattini:2003wp,Becattini:2005xt}.
On the other hand, hadronic cascade models failed to describe 
such structures in the ratios%
~\cite{Afanasiev:2002mx,Alt:2007aa}
even though some models include collective effects
such as string fusion to color ropes
\footnote{The color rope formation scenario seems to describe the
$K^+/\pi^+$ ratio up to the maximum of the horn, but it overestimates
above the energies at the maximum.}.
However, recently, the parton-hadron-string dynamics (PHSD) transport model
reproduced horn structure
by the interplay between the effects of chiral symmetry restoration 
and deconfinement into quark-gluon degrees of freedom~\cite{Cassing:2015owa}.
This is the first explanation of the horn from a dynamical approach.
In this paper, we shall show that our dynamical approach
also describes well the excitation function of the $K^+/\pi^+$ ratio
by taking into account an incomplete thermalization of the system.

This paper is organized as follows:
Section~\ref{sec:model} describes our transport approach.
Section~\ref{sec:result} presents the results for the excitation
function of the mean transverse mass, multiplicities, and
particles ratios.
The conclusions are given in Sec.~\ref{sec:conclusion}.

\section{Dynamical model}
\label{sec:model}

We use the JAM transport model~\cite{JAMorg}
for the description of nonequilibrium dynamics,
which is based on the particle degrees of freedom:
hadrons and strings.
JAM describes the time evolution of the phase space of $N$ particles
by the so-called cascade method.
Particle production in JAM is modeled by the excitation and
decay of resonances and strings as employed by other transport models
\cite{RQMD1995,UrQMD1,UrQMD2}.
We use the same string excitation scheme as the HIJING model~\cite{HIJING},
and the Lund model for string fragmentations in PYTHIA6~\cite{pythia6}.
Secondary products from decays of resonances or strings
can interact with each other via binary collisions.
A detailed description of the hadronic cross sections
and cascade method implemented in the JAM model can be found
in Ref.~\cite{JAMorg,Hirano:2012yy}.
In this work, we include neither hadronic mean-field nor modified scattering style
in two-body collisions~\cite{Isse,Nara:2016phs,Nara:2016hbg,Nara:2017qcg},
thus particle trajectories in the JAM cascade are straight lines
until particles scatter with each other or decay.

We consider a dynamical coupling of the microscopic transport model
and macroscopic hydrodynamics. Specifically,
the JAM model is coupled with
hydrodynamics by source terms $J^\mu$ and $\rho$:
\begin{equation}
 \partial_\mu T_f^{\mu\nu} = J^\nu, ~~ \partial_\mu N_f^\mu=\rho,
\end{equation}
where $T_f^{\mu\nu}$ is the energy-momentum tensor of the fluids.
We assume the ideal fluid $T_f^{\mu\nu}=(e+p)u^\mu u^\nu - pg^{\mu\nu}$,
where $u^\mu$ is the four-fluid velocity, $e$ and $p$ are the local energy
density and pressure.
Here $N_f^\mu=n u^\mu$ is the baryon current.

In our practical implementation of the source term $J^\mu$ and $\rho$,
particles are converted into fluid elements when they decay from
strings or hadronic resonances and if the local energy density
(the sum of the contributions from particles and fluids)
at the point of decay exceeds the fluidization energy density $e_\text{f}$
in order to ensure the core-corona separation.
We use $e_\text{f}=0.5$ GeV/fm$^3$ as a default value, which is the
same as the particlization energy density $e_\text{p}$ introduced later.

One expects that the fluidization energy density $e_\text{f}$
must be the same as the particlization energy density $e_\text{f}$
in the static equilibrium state.
However, $e_\text{f}$ may be different from $e_\text{p}$
at a highly nonequilibrium state such as occur in the initial stages of
heavy-ion collisions.
Note that the fluidization condition 
also depends on the baryon density.
The transition energy density to QGP at high baryon densities
would be higher than the largest value of 0.5 GeV/fm$^3$ that is
obtained by the lattice QCD calculations at vanishing chemical potentials
\cite{Bazavov:2014pvz}.
Thus we do not exclude possibilities of higher values of
$e_\text{f}=0.8$ and 1.0 GeV/fm$^3$.

In the case of string decay,
decay products are absorbed into fluids after their formation times
with the same criterion as above.
Leading particles, which have original constituent quarks
from string decay, are not converted into fluids,
in order to keep
the same baryon stopping power as in the JAM cascade model.
Thus, our source terms in which particles are entirely absorbed
into fluid elements within a time step $\Delta t$ take the form
\begin{align}
 J^\mu(\bm{r}) &= \frac{1}{\Delta t} \sum_i p^\mu_i(t) G(\bm{r}-\bm{r}_i(t)), \\
 \rho(\bm{r}) &= \frac{1}{\Delta t} \sum_i B_i G(\bm{r}-\bm{r}_i(t)),
\end{align}
where $B_i$ is the baryon number of $i$-th particle,
and the sum runs over the particles to be absorbed into fluid elements
at the time interval between $t$ and $t+\Delta t$.
The Gaussian smearing profile is given by
\begin{equation}
 G(\bm{r})=\frac{\gamma}{(2\pi\sigma^2)^{3/2}}\exp\left(
      -\frac{\bm{r}^2 + (\bm{r}\cdot\bm{u})^2}{2\sigma^2}
       \right),
\label{eq:gauss}
\end{equation}
where $\bm{u}=\bm{p}/m$ is the
velocity of the particle,
and $\gamma=p^0/m$~\cite{Oliinychenko:2015lva},
in which the profile is Lorentz contracted to ensure the Lorentz invariance
in the combination of $G(\bm{r}) d^3r$.
In this work, we use $\sigma=0.5$ fm.

As we force the thermalization of the system
by hand in our approach neglecting viscous effects,
there is no way to match all the quantities
between particles and fluids. Therefore, we simply assume that
the local energy density of the particles 
is obtained from $T_p^{\mu\nu}$ in the Eckart frame defined
by $N_p^\mu$, where the particle currents are calculated  as
\begin{align}
 T_p^{\mu\nu}(\bm{r}) &=\sum_i
     \frac{p_i^\mu(t) p_i^\nu(t)}{p^0_i(t)}G(\bm{r}-\bm{r}_i(t)),
     \label{eq:gaussp}\\
 N_p^{\mu}(\bm{r}) &=\sum_i
     \frac{p_i^\mu(t)}{p^0_i(t)}G(\bm{r}-\bm{r}_i(t)),
\end{align}
where the summation runs over all particles,
and $p_i^\nu(t)$ and $\bm{x}_i(t)$ are the four-momenta and the coordinates
of the $i$-th particle, respectively.
We have checked that our final results are unaffected
when we instead reconstruct the local energy density
from the computational-frame particle energy and momentum using the EoS.

Hydrodynamical equations are solved numerically by employing 
the Harten–Lax–van Leer–Einfeldt (HLLE) algorithm%
~\cite{Schneider:1993gd,Rischke1,Karpenko:2013wva}
in three spatial dimensions with operator splitting method.
There are some modifications from the original implementations%
~\cite{Schneider:1993gd,Rischke1,Karpenko:2013wva}.
First, cell interface values $Q_\pm=(\bm{v}_\pm, e_\pm, p_\pm, n_{\pm})$
for the local variables of fluid velocity $\bm{v}$,
energy density $e$ and baryon density $n$ 
are obtained via the monotonized central-difference (MC) limiter%
~\cite{VanLeer,Okamoto:2017ukz}
instead of using minmod slope limiter.
In addition, we take space averages
of the linear interpolation function for the cell interface values
at each cell boundary.
Then, we construct cell interface values for the conserved quantities
\begin{align}
 T^{00}_{\pm} &= (e_{\pm}+p_{\pm})\gamma^2_{\pm} - p_{\pm}, \\
 T^{0i}_{\pm} &= (e_{\pm}+p_{\pm})\gamma^2_{\pm} v^i_{\pm},~(i=x,y,z)\\
 N^0_{\pm}  &= \gamma_\pm n_{\pm},
\end{align}
from which we compute numerical flux by using the HLLE algorithm.
The cell size of $\Delta x=\Delta y=\Delta z=0.3$ fm,
and the time-step size of $\Delta t=0.15$ fm/$c$ are used in the present
study.

EoS which covers all baryon chemical potentials $\mu_B$
in the QCD phase diagram has not yet been available
from lattice QCD calculations. 
In lattice QCD, EoS at finite baryon densities is usually obtained
by the Taylor expansion of the pressure in $\mu_B/T$ around $\mu_B=0$,
and it is extended up to $\mu_B \approx 300-400$ MeV~\cite{Bazavov:2017dus}.
Since it does not cover all the baryon chemical potential needed
for our beam energy regions,
we take EoS from phenomenological model calculations in this work.
We employ an equation of state, EOS-Q~\cite{EOSQ}, which
exhibits a first-order phase transition  between massless quark-gluon
phase with bag constant $B^{1/4}=235$ MeV and hadronic gas with
resonances up to 2 GeV.
In the hadronic phase, we include a baryon density dependent
single-particle repulsive potential $V(\rho_B)=K\rho_B$
with $K=0.45$ GeVfm$^3$ for baryons.
In the present work, all results are obtained by using EOS-Q.
We will report EoS dependence on the particle productions in detail elsewhere.

Fluid elements are converted into particles by using the positive part
of the Cooper-Frye formula~\cite{Cooper:1974mv}
\begin{equation}
\Delta N_i=\frac{g_i}{(2\pi)^3} \int 
  \frac{d^3p}{E}
  \frac{[\Delta\sigma\cdot p]_+}
{\exp[(p\cdot u-\mu_i)/T]\pm1},
\end{equation}
where $[\cdots]_+=\theta(\cdots)|\cdots|$,
$\Delta\sigma_\mu$ is the hypersurface element,
and $g_i$ and $\mu_i$ are the spin degeneracy factor
and the chemical potential
for $i$th hadron species, respectively.
CORNELIUS 1.4 is used to compute freeze-out hypersurface%
~\cite{Huovinen:2012is}.
We use a method similar to that of Ref.~\cite{Pratt:2014vja}
for the Monte-Carlo sampling of particles.
When potentials are included in the EoS,
we use the effective baryon chemical potential
in the Cooper-Frye formula
\begin{equation}
\mu^\text{eff}_B= \mu_B -V(\rho_B)=\mu_B-K\rho_B
\label{eq:mueff}
\end{equation}
to ensure the smooth transition from fluids to particles.
We assume that particlization occurs at the energy density
of $e_\text{p}=0.5$ GeV/fm$^3$ in the present work.

%%%%%%%%%%%%%%%%%%%%%%%%%%%%%%%%%%%%%%%%%%%%%%%%%%%%%%%%%
\begin{figure}[tbh]
\includegraphics[width=8.0cm]{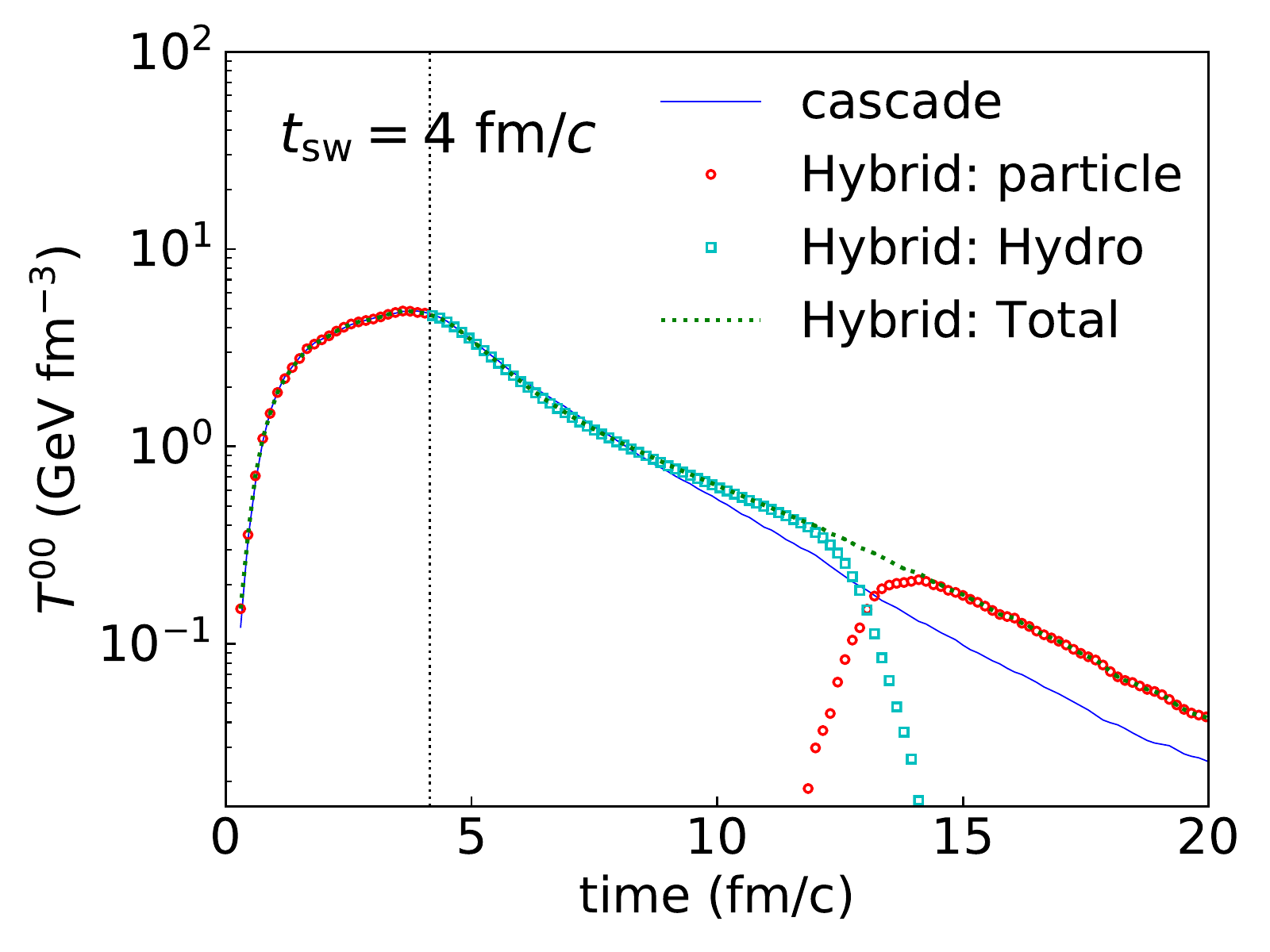}
\includegraphics[width=8.0cm]{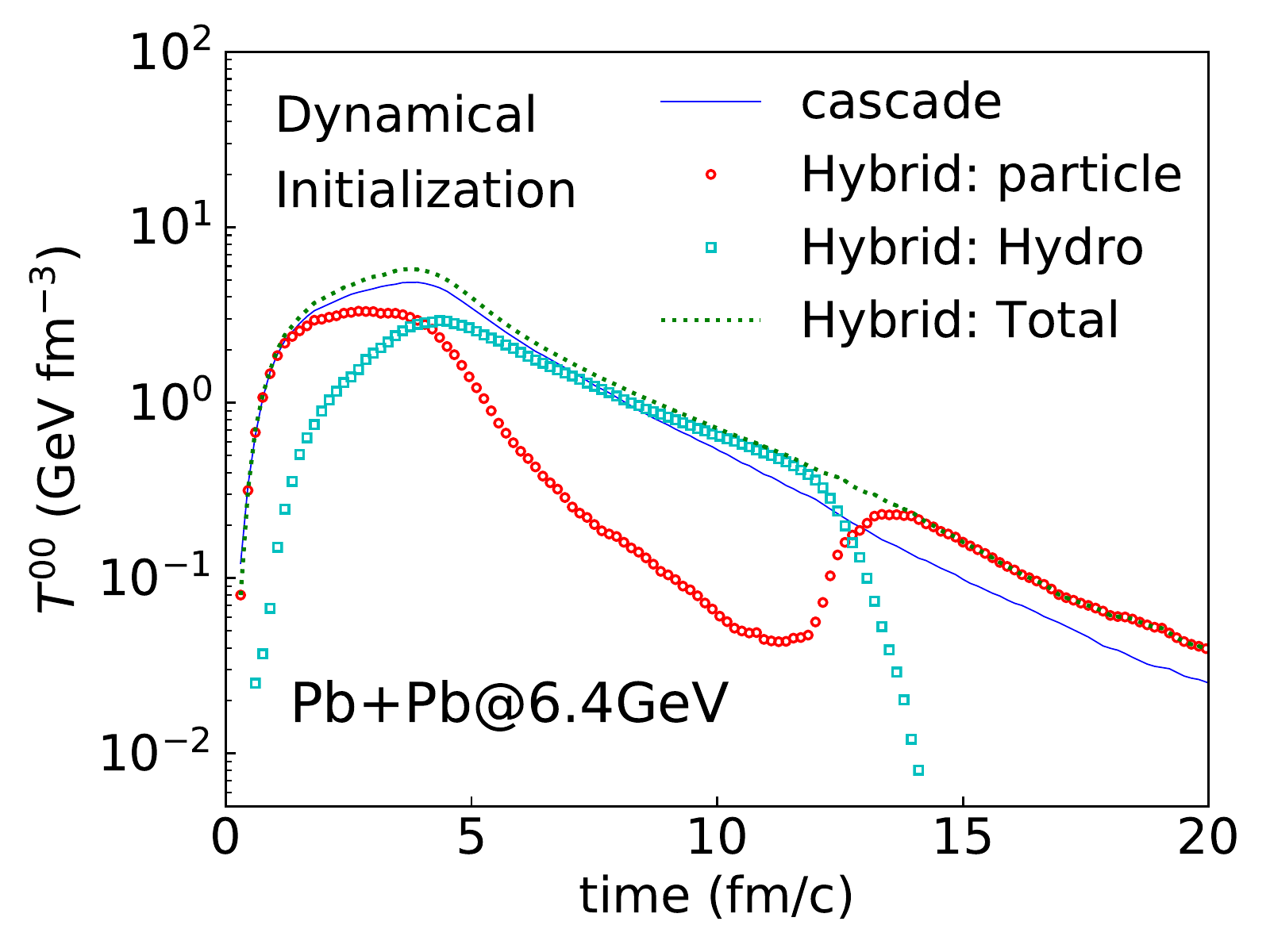}
\caption{Time evolution of the energy density at the coordinate origin
$(x,y,z)=(0,0,0)$ from JAM hybrid simulations
with $e_\text{f}=e_\text{p}=0.5$ GeV/fm$^3$
in central Pb + Pb collisions
at $\srtNN=6.4$ GeV ($E_\text{lab}=20A$ GeV).
Upper panel shows the results from calculations by the fixed switching time
at $t_\text{sw}=4$ fm/$c$.
The result from a dynamical initialization is shown in the lower panel.
The circles show the contribution from particles,
and the squares show the fluid contribution to the energy density.
}
\label{fig:evol20}
\end{figure}
%%%%%%%%%%%%%%%%%%%%%%%%%%%%%%%%%%%%%%%%%%%%%%%%%%%%%%%%%

%\subsection{Time evolution}
Let us first examine the time evolutions of the system.
In Fig.~\ref{fig:evol20}, the time evolutions of
the energy density $T^{00}$ at the coordinate origin of
central Pb + Pb collisions ($b<3.4$ fm) at 
$\srtNN=6.4$ GeV
($E_\text{lab}=20A$ GeV)
are shown.
The average is taken over about 1000 events.
In the upper panel of Fig.~\ref{fig:evol20}, 
the results from the hybrid simulations with a fixed switching time
of $t_\text{sw}=4$ fm/$c$ are displayed,
where the switching time
$t_\text{sw}=2R/(\gamma v)$ is assumed by the condition that
hydrodynamical evolution starts immediately after
the two nuclei have passed through each other,
where $R$ and $v$ are the radius and the incident velocity of
the colliding nuclei, respectively, and $\gamma$ is the Lorentz factor.
The circles depict the energy density evolution from particles in JAM,
while the squares correspond to the energy density of hydrodynamics.
The dotted line is the sum of two contributions.
The energy density of the particle contribution is computed by using
the Gaussian smearing profile Eq.~(\ref{eq:gaussp})
at the coordinate origin $(x,y,z)=(0,0,0)$, 
and the energy density of the fluids
corresponds to the value of the cell at the origin $T^{00}(0,0,0)$.
Until the switching time $t_\text{sw}=4$ fm/$c$, time evolution
of the energy density is identical to the cascade simulation (solid line)
as it should be.
After switching to fluids,
the expansion of the system becomes
slower than that in the cascade simulation.
We have found that
it is important to include the effects of potential in the Cooper-Frye formula
by using e.g., the effective baryon chemical potential~Eq.~(\ref{eq:mueff})
in order to ensure the smooth transition from
fluids to particles at the particlization,
in case mean-field potential is included in the EoS.

The lower panel of Fig.~\ref{fig:evol20} shows the time evolution
of energy density in the case of dynamical initialization.
The energy density of particles reaches the maximum
value at $t=2.5$ fm/$c$, and 
the energy density of the fluids gradually
increases up to 3 GeV/fm$^3$ at 5 fm/$c$.
It is also important to observe that hydrodynamical evolution already starts
before two nuclei pass through each other.
The sum of energy densities from particles and fluid elements
is shown by the dotted line and is larger than the cascade result.
The dynamical initialization leads to a very different dynamical
evolution of the system compared wiht the simulation
with a fixed switching time.

%%%%%%%%%%%%%%%%%%%%%%%%%%%%%%%%%%%%%%%%%%%%%%%%%%%%%%%%%
\begin{figure}[tbh] \includegraphics[width=8.0cm]{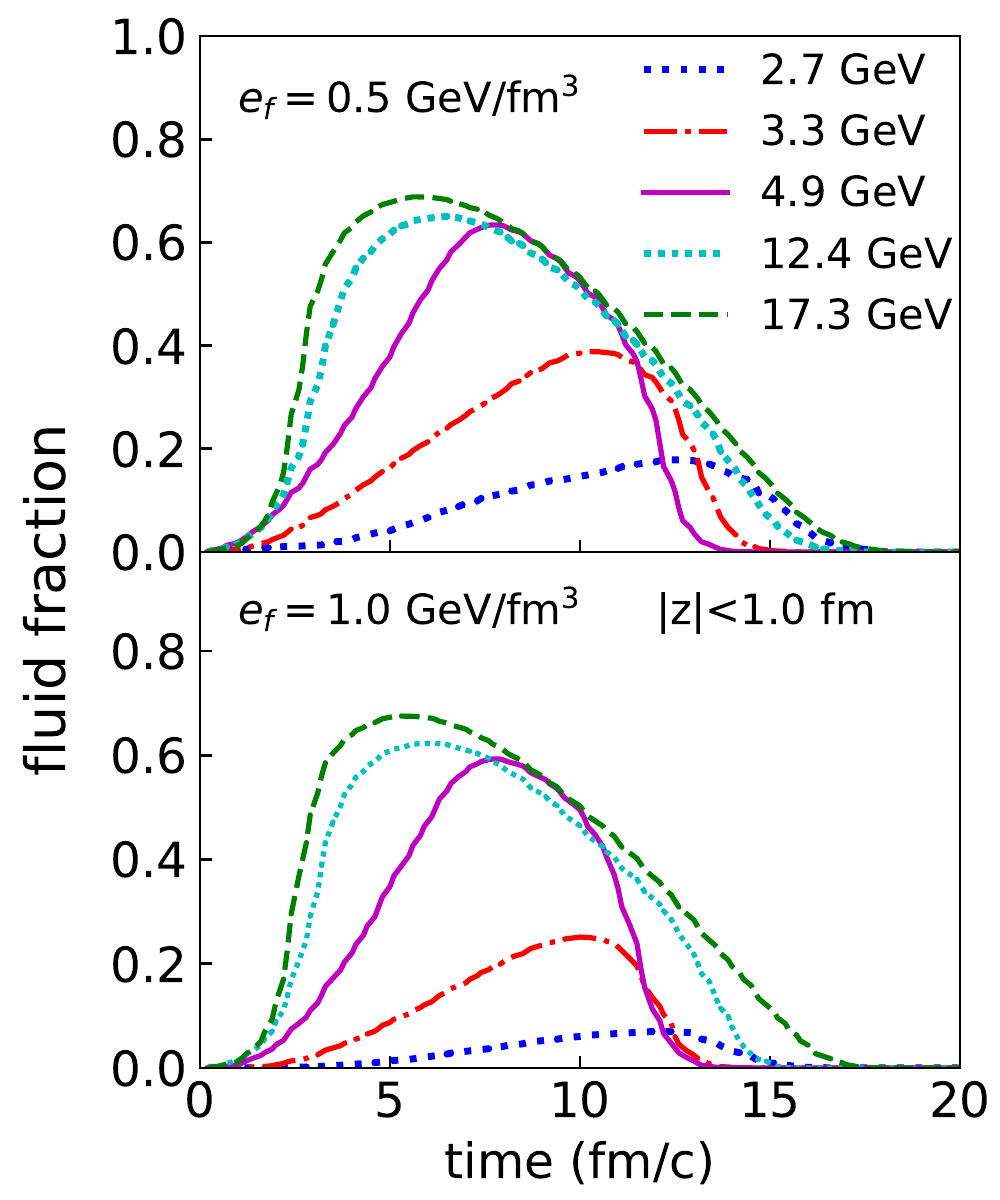}
\caption{Time evolution of the fractions of the fluid energy for
$e_\text{f}=0.5$ GeV/fm$^3$ (upper panel) and $e_\text{f}=1.0$ GeV/fm$^3$
(lower panel) at the central region $|z|<1.0$ fm from JAM hybrid simulations
in central Au + Au at $\srtNN=2.7, 3.3$, and 4.9 GeV,
and Pb + Pb collisions at $\srtNN=12.4$ and 17.3 GeV.
%which correspond to $E_\text{lab}=2, 4, 10.7, 80$, and 158$A$ GeV.
}
\label{fig:fluidfrac} \end{figure}
%%%%%%%%%%%%%%%%%%%%%%%%%%%%%%%%%%%%%%%%%%%%%%%%%%%%%%%%%
Figure~\ref{fig:fluidfrac} shows the time evolution of the fraction of the
fluid energy in the central region of $|z|<1$ fm, in which the energy density
of the fluid elements is greater than 0.5 GeV/fm$^3$
for central Au + Au and Pb + Pb
collisions at $\srtNN=2.7, 3.3, 4.9, 12.4$, and 17.3 GeV.
%$E_\text{lab}=2, 4, 10$, and 158$A$ GeV.
The results for the fluidization energy densities $e_\text{f}=0.5$ GeV/fm$^3$
and 1.0 GeV/fm$^3$ are shown in the upper panel and the lower panel of
Fig.~\ref{fig:fluidfrac}, respectively.  The fluid fraction increases slowly
with time at lower beam energies.  As beam energy becomes higher, the fluid
fraction increases rapidly, and particle-fluid conversion process becomes
close to that in the single thermalization time simulations.  The fluid fraction is
insensitive to the $e_\text{f}$ value for high beam energies, while it is
sensitive at lower energies.  The fluid fraction at $\srtNN=17.3$ GeV reaches
about 70\% at $t=6$ fm/$c$, which is smaller than the UrQMD hybrid core-corona
model~\cite{Steinheimer:2011mp}, in which almost the entire system enters the
hydrodynamics at the highest SPS energies in central Pb + Pb collisions.
One of the main reasons is that, in our approach, preformed hadrons
(hadrons within their formation times)
and leading hadrons are not
converted into fluids, while in the UrQMD hybrid model, they are included in
the hydrodynamical evolution at the switching time, which is necessary for the
total energy-momentum conservation.  
Additionally, in this work, we assume that
preformed hadrons are converted into fluids after their formation time,
which implies that
the formation time of hadrons equals the local thermalization time of the
system at dense region.  Nevertheless, it would be interesting to investigate
the influence of shorter thermalization times on the dynamics in our approach.

\section{Results}
\label{sec:result}

We compare our results from JAM+hydro hybrid model
with the results from JAM cascade model
and the experimental data in central Au+Au/Pb + Pb collisions.
We perform calculations with the impact parameter range $b<4.0$ fm
for 7\% central Pb + Pb at $\srtNN=$6.4--12.4 ($E_\text{lab}=20$--$80A$ GeV),
and $b<3.4$ fm for 5\% central collision for the other collisions.

%%%%%%%%%%%%%%%%%%%%%%%%%%%%%%%%%%%%%%%%%%%%%%%%%%%%%%%%%
\begin{figure}[tbh]
\includegraphics[width=9.0cm]{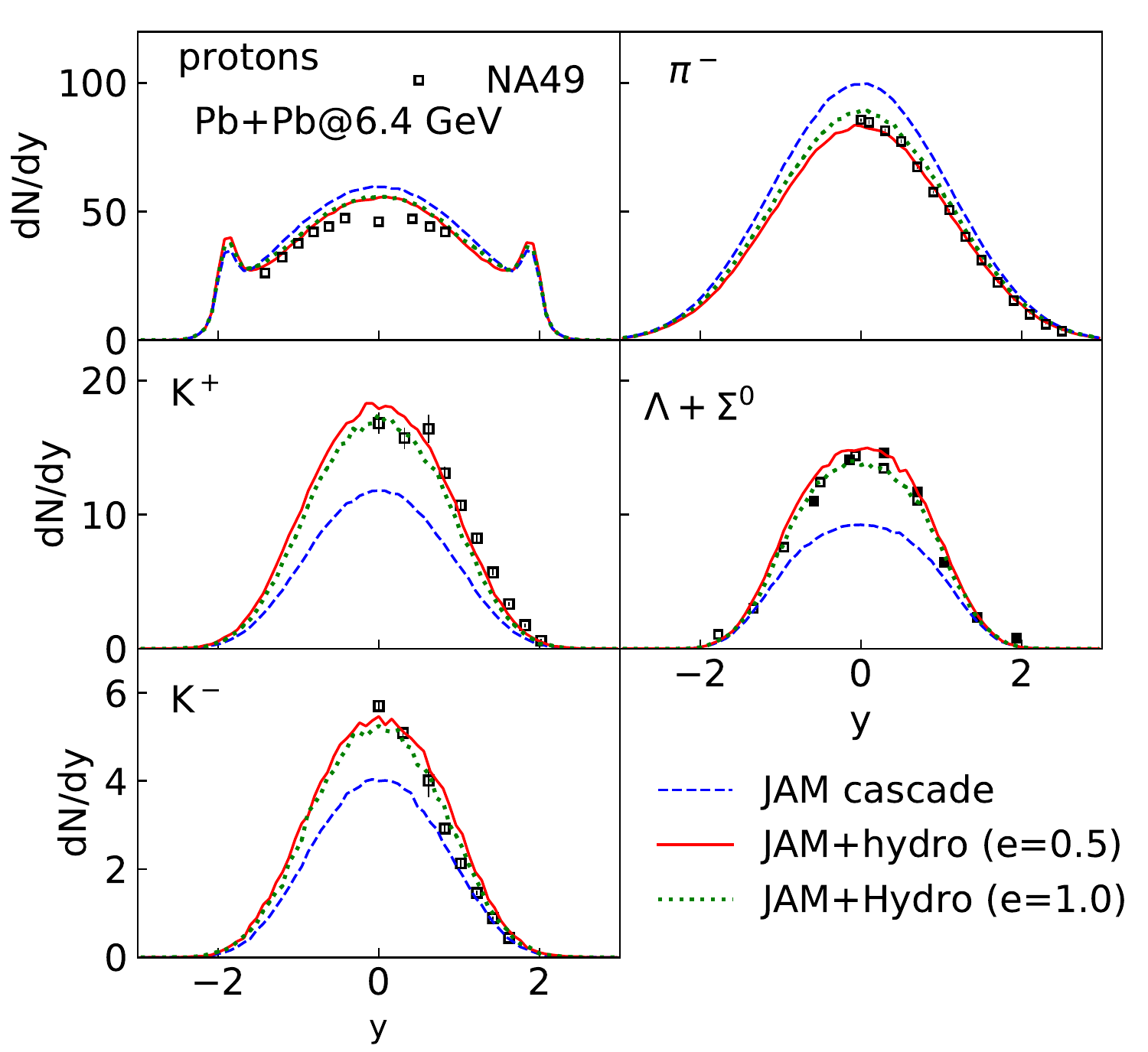}
\caption{Rapidity distributions of protons, negatively charged pions,
negatively and positively charged kaons, and $\Lambda$s
in the central Pb + Pb collisions at $\srtNN=6.4$ GeV.
The results from JAM cascade (dashed lines) and JAM + hydro calculations
with $e_\text{f}=0.5$ (dotted lines) and 1.0 GeV/fm$^3$ (solid lines)
are compared with the experimental data~\cite{Alt:2007aa,Alt:2008qm,Blume:2007kw}.
}
\label{fig:dndy20}
\end{figure}
%%%%%%%%%%%%%%%%%%%%%%%%%%%%%%%%%%%%%%%%%%%%%%%%%%%%%%%%%
First we compare the results of our approach in central Pb + Pb collisions
at $\srtNN=6.4$ GeV for the rapidity (Fig.~\ref{fig:dndy20})
%and transverse mass distributions (Fig.~\ref{fig:dndmt20})
with the NA49 data
~\cite{Alt:2007aa,Alt:2008qm,Blume:2007kw}
to see the effects of hydrodynamical evolution.
We also compare the results from different values of
the fluidization energy densities $e_\text{f}=0.5$ and 1.0 GeV/fm$^3$.
In the hybrid simulations, the rapidity distribution of protons is
slightly lower than those in the cascade simulations, and
stopping power of two nuclei is similar between
hybrid and cascade simulations for both $e_\text{f}=0.5$ and 1.0 GeV/fm$^3$.
This is the consequence of our implementation of the model, in which
leading hadrons from string fragmentation are not converted
into fluid elements.
However, there are big differences in the yields of pions, kaons, and $\Lambda$s.
Pion yields are suppressed by the hydrodynamical evolution, while strange hadrons
are enhanced compared with the cascade simulation results.
When one uses a larger value of the fluidization energy density
$e_\text{f}=1.0$ GeV/fm$^3$, there are a slight increase of pion yields, 
and slight decreases of kaon and $\Lambda$ yields,
thus the sensitivity of the value of the fluidization energy density is small.
We also note that a lower particlization energy density
$e_\text{p}=0.3$ GeV/fm$^3$ yields almost the same results as
compared with our default value $e_\text{p}=0.5$ GeV/fm$^3$.

%%%%%%%%%%%%%%%%%%%%%%%%%%%%%%%%%%%%%%%%%%%%%%%%%%%%%%%%%
\begin{figure}[tbh]
\includegraphics[width=9.0cm]{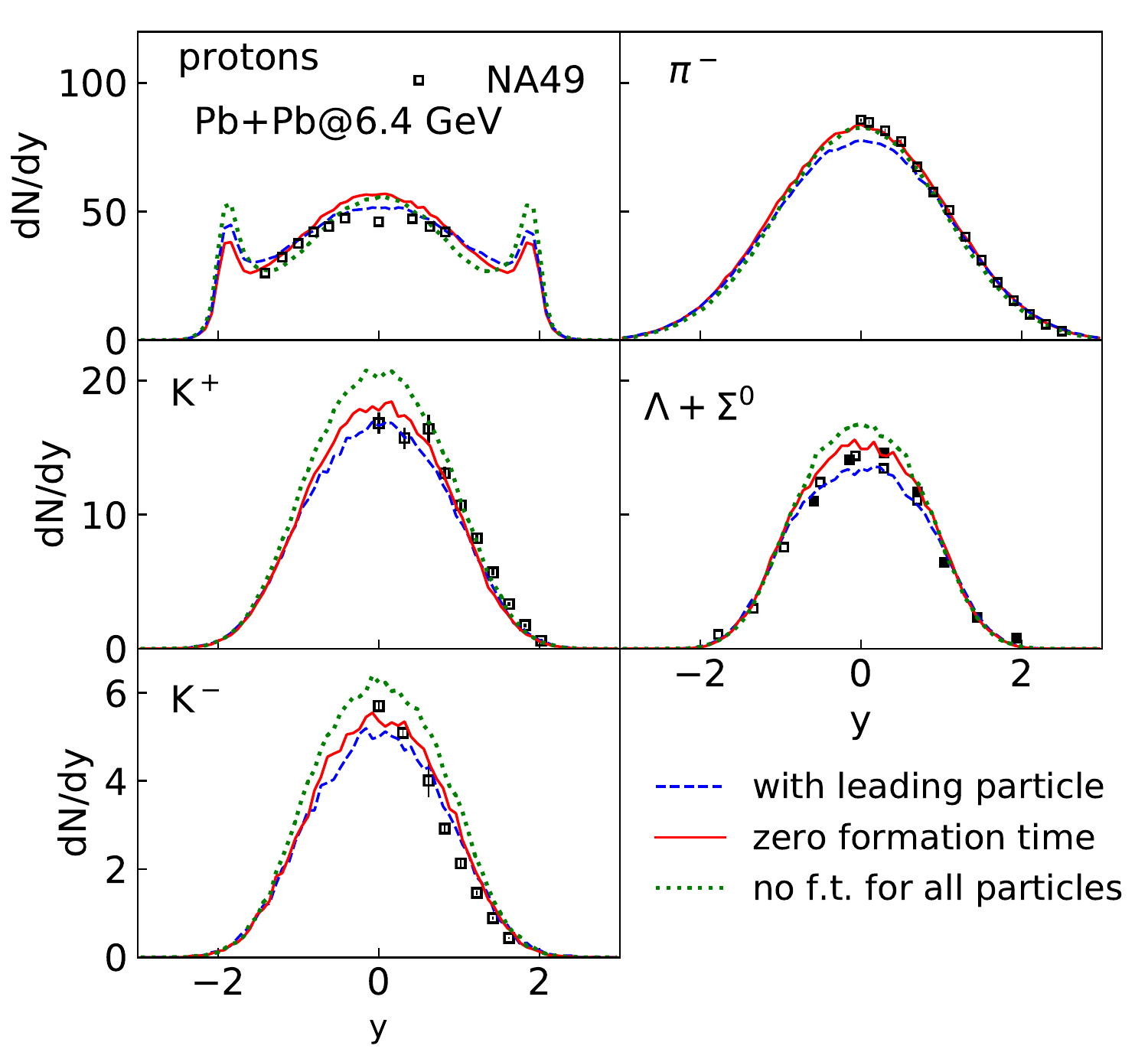}
\caption{Same as Fig.~\ref{fig:dndy20}, but
the results from JAM + hydro simulations in which leading particles are also
converted into fluid after their formation times (dashed lines),
the results from the calculations without formation times
but leading particles are not converted into fluid (solid lines),
and the results from the calculations without formation times
for all particles including leading particles.
}
\label{fig:dndy20c}
\end{figure}
%%%%%%%%%%%%%%%%%%%%%%%%%%%%%%%%%%%%%%%%%%%%%%%%%%%%%%%%%
In our default implementation, leading particles are not converted into
fluid, and the other particles are converted into fluid after their formation
times. To see the effects of our model assumptions,
we have performed several calculations with different implementations.
In Fig.~\ref{fig:dndy20c}, the results of the simulations in which
the leading particles are also incorporated in the fluid after their
formation times are shown.
It is seen that baryon stopping power becomes less, and particle yields
are also somewhat less than the default calculations.
This may be because we have less initial hard collisions which are responsible
to the particle productions and rapidity loss of leading hadrons.
To see the effects of the formation time,
we plot the results of calculations in which newly produced particles
are converted into fluid with zero formation time in Fig.~\ref{fig:dndy20c}.
The results show that there are no strong sensitivities
of the particle productions to the formation times.

However, as depicted by the dotted lines in Fig.~\ref{fig:dndy20c},
when all particles are converted into fluid without formation times,
we see the overprediction of strange particles and stronger proton stopping,
since this implementation has similar effects as the one-fluid simulations.
In any cases, particle production is not strongly affected by the
details of implementations.

%%%%%%%%%%%%%%%%%%%%%%%%%%%%%%%%%%%%%%%%%%%%%%%%%%%%%%%%%
\begin{figure}[tbh]
\includegraphics[width=8.5cm]{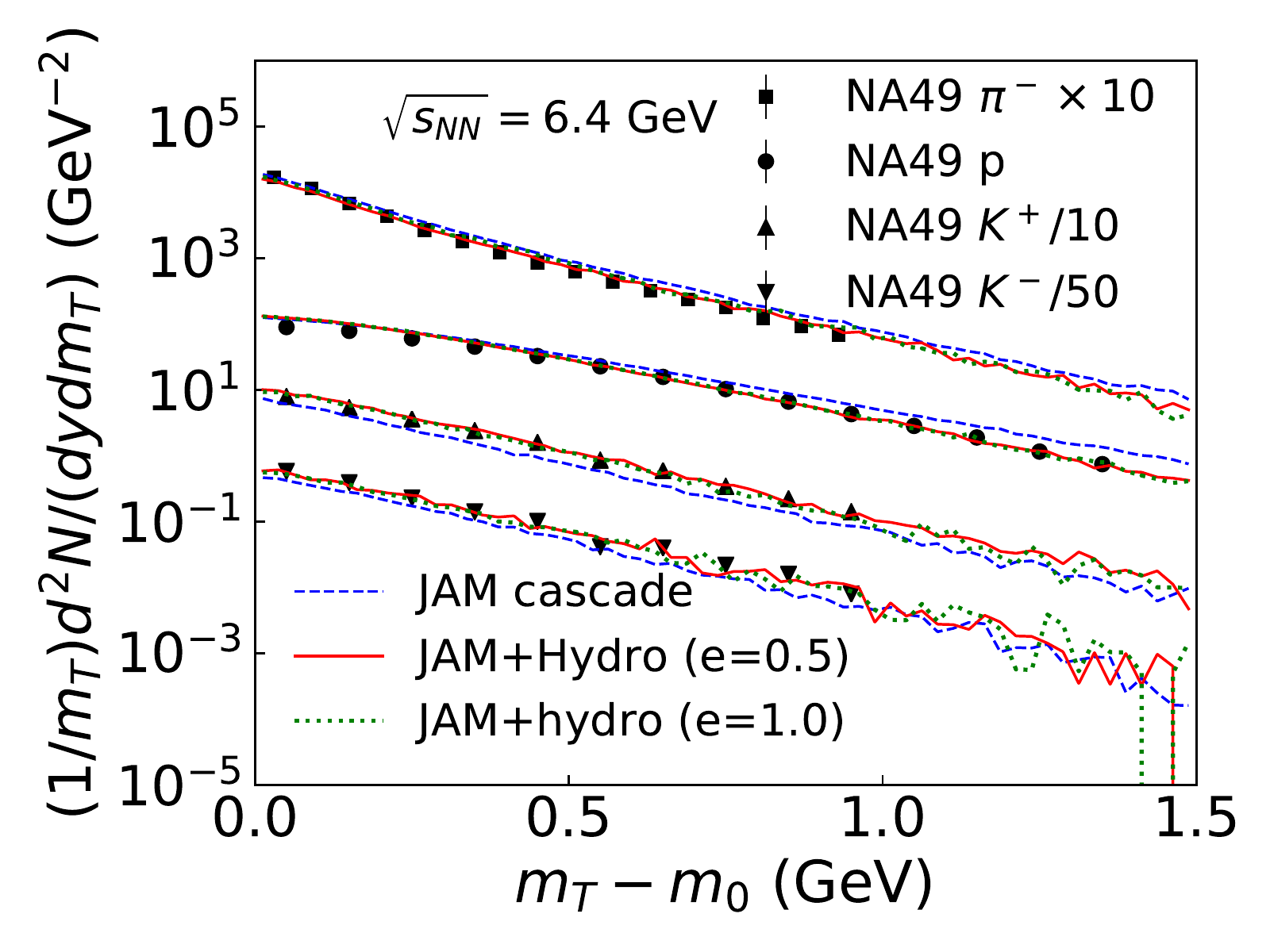}
\caption{Transverse mass distributions of protons, negative pions,
positive and negative kaons
in central Pb + Pb collisions at $\srtNN=6.4$ GeV.
The results from JAM cascade (dashed lines) and JAM + hydro calculations
with $e_\text{f}=0.5$ (solid lines) and 1.0 GeV/fm$^3$ (dotted lines)
are compared with experimental data%
~\cite{Alt:2007aa,Alt:2006dk}.
}
\label{fig:dndmt20}
\end{figure}
%%%%%%%%%%%%%%%%%%%%%%%%%%%%%%%%%%%%%%%%%%%%%%%%%%%%%%%%%

In Fig.~\ref{fig:dndmt20}, the transverse mass distributions of identified
particles are depicted.
The hybrid simulation exhibits similar slopes
for the transverse mass distributions of pions and kaons
to cascade simulations, and they are in good agreement with the data.
It is also observed that the differences with respect to the parameter
of the fluidization energy density are practically invisible.
The harder proton slopes than the experimental data
predicted by nonequilibrium transport approach in the JAM cascade model
is improved by the hybrid model calculations.

%%%%%%%%%%%%%%%%%%%%%%%%%%%%%%%%%%%%%%%%%%%%%%%%%%%%%%%%%
\begin{figure}[tbh]
\includegraphics[width=8.0cm]{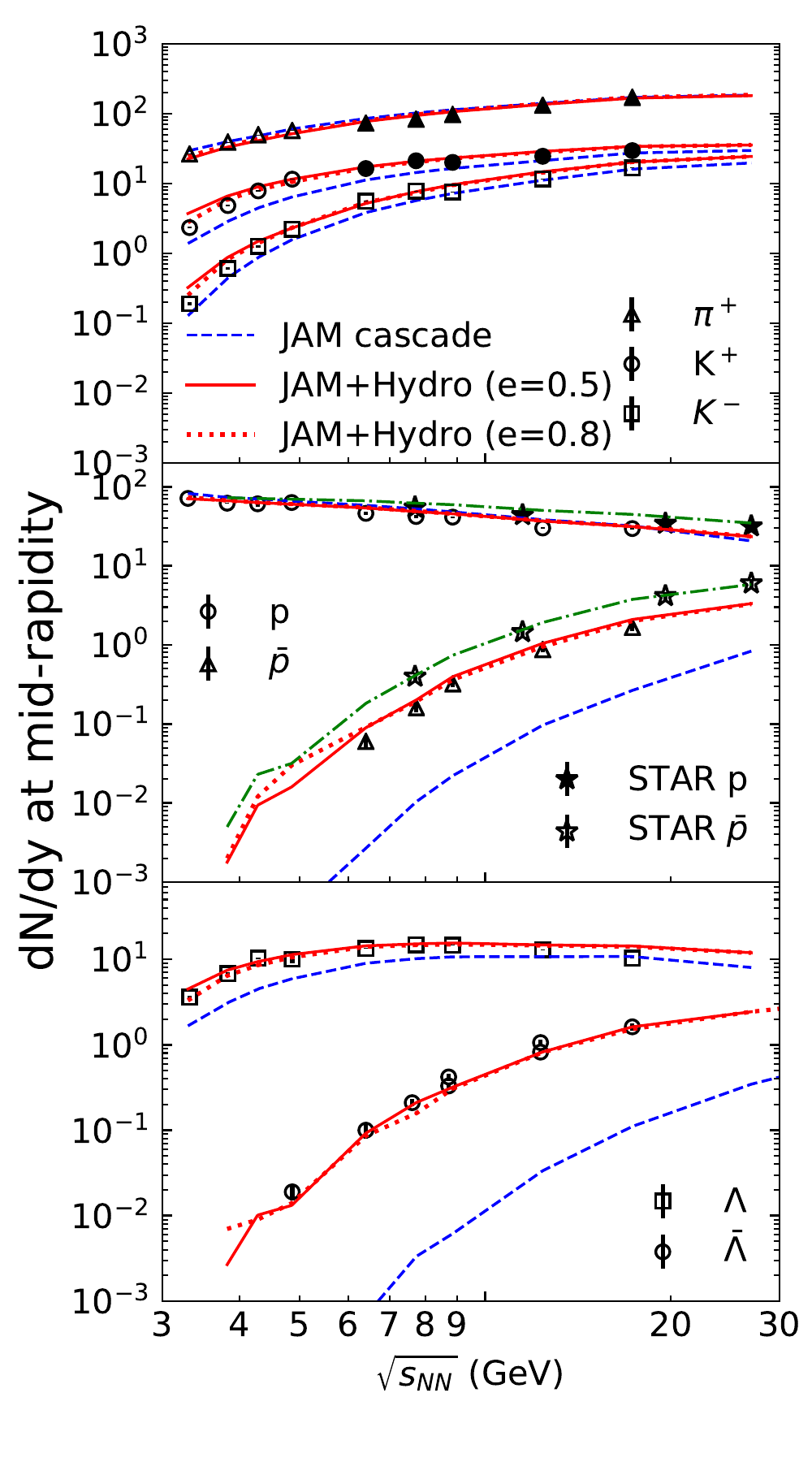}
\caption{Beam energy dependence of particle multiplicities at midrapidity
in central Au + Au and Pb + Pb collisions.
The experimental data are taken from
Ref.~\cite{Ahle:1999uy,NA49compilation,Blume:2011sb,
Ahle:1998jc,Klay:2001tf}.
JAM+hydro results for (anti-)protons with weak-decay contributions
(dashed-dotted lines)
are compared with the STAR data which contain
the weak-decay feed-down contributions to (anti)-proton yield. 
}
\label{fig:mult}
\end{figure}
%%%%%%%%%%%%%%%%%%%%%%%%%%%%%%%%%%%%%%%%%%%%%%%%%%%%%%%%%
Figure~\ref{fig:mult} shows the beam-energy dependence of
the particle yields $dN/dy$ at midrapidity $|y|<0.5$
for positively charged pions and kaons,
negatively charged kaons, protons, antiprotons, $\Lambda$s,
and anti-$\Lambda$s
from the cascade and the hybrid simulations.
JAM cascade model slightly overestimates pion yields 
at $\srtNN<10$ GeV, and underestimates strange particles.
It is known that most of the standard microscopic transport models
overestimate pion yields%
~\cite{Weber:2002pk,Wagner:2004ee,Konchakovski:2014gda}.
Here the JAM+hydro hybrid approach improves this situation:
hybrid calculations suppress pion yields and enhance strange particles,
and good agreement with the data is obtained.
It should be emphasized that the antibaryon productions
such as antiprotons and anti-$\Lambda$s are also significantly improved,
and good agreement with the data is obtained by the hybrid model.
We have checked the dependence of the fluidization energy density
$e_\text{f}$ on the multiplicities.
The results from a larger value of $e_\text{f}=0.8$ GeV/fm$^3$
relative to the value of $e_\text{f}=0.5$ GeV/fm$^3$
yield less strange particles at lower beam energies, but
its influence on the strange particle yields
is very small at higher beam energies.

%%%%%%%%%%%%%%%%%%%%%%%%%%%%%%%%%%%%%%%%%%%%%%%%%%%%%%%%%
\begin{figure}[tbh]
\includegraphics[width=8.0cm]{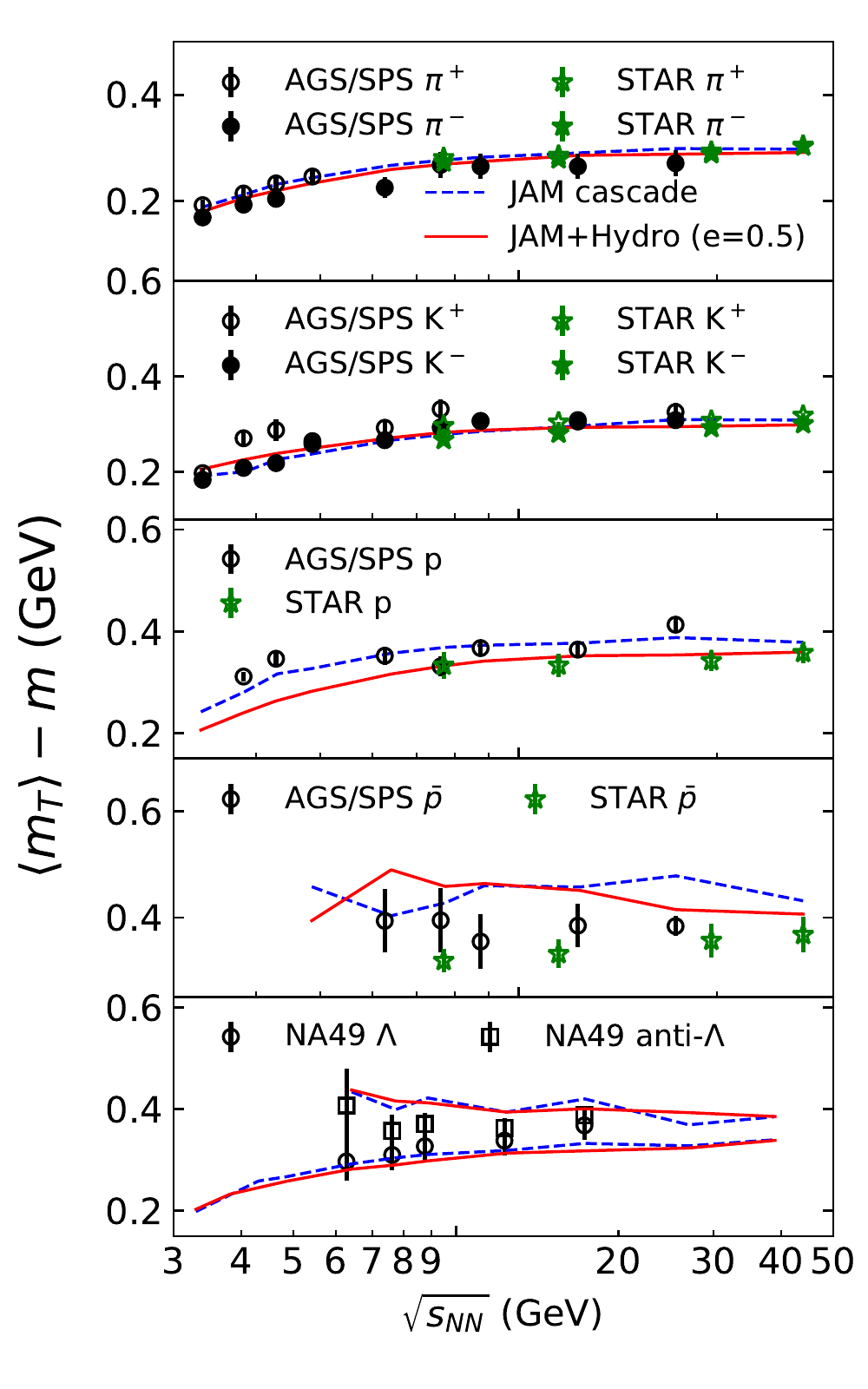}
\caption{Beam-energy dependence of mean transverse mass
for pions, kaons, protons, antiprotons, $\Lambda$s, and anti-$\Lambda$s
at midrapidity
in central Au + Au/Pb + Pb collisions.
The experimental data are taken from Ref.%
~\cite{Adamczyk:2017iwn,Abelev:2009bw,Alt:2008qm,Alt:2006dk}.
}
\label{fig:mt}
\end{figure}
%%%%%%%%%%%%%%%%%%%%%%%%%%%%%%%%%%%%%%%%%%%%%%%%%%%%%%%%%

Figure~\ref{fig:mt} depicts the beam-energy dependence of the mean
transverse mass for pions, kaons, protons, antiprotons, $\Lambda$s,
and $\bar{\Lambda}$s compared with the experimental data%
~\cite{Adamczyk:2017iwn,Abelev:2009bw,Alt:2008qm,Alt:2006dk}.
We observe that
the mean transverse mass in cascade
is essentially unaffected by the hydrodynamics except protons,
and models reproduce the experimental data of pions and kaons very well.
The JAM cascade approach describes the proton slopes better 
at lower beam energies $\srtNN\leq5$ GeV compared with the hybrid approach,
while the hybrid approach describes proton slopes better at higher energies.
We do not show the results with $e_\text{f}=1.0$ GeV/fm$^3$,
because the dependence of the mean transverse mass spectra
on the fluidization energy densities
is very small.

%%%%%%%%%%%%%%%%%%%%%%%%%%%%%%%%%%%%%%%%%%%%%%%%%%%%%%%%%
\begin{figure}[tbh]
\includegraphics[width=8.0cm]{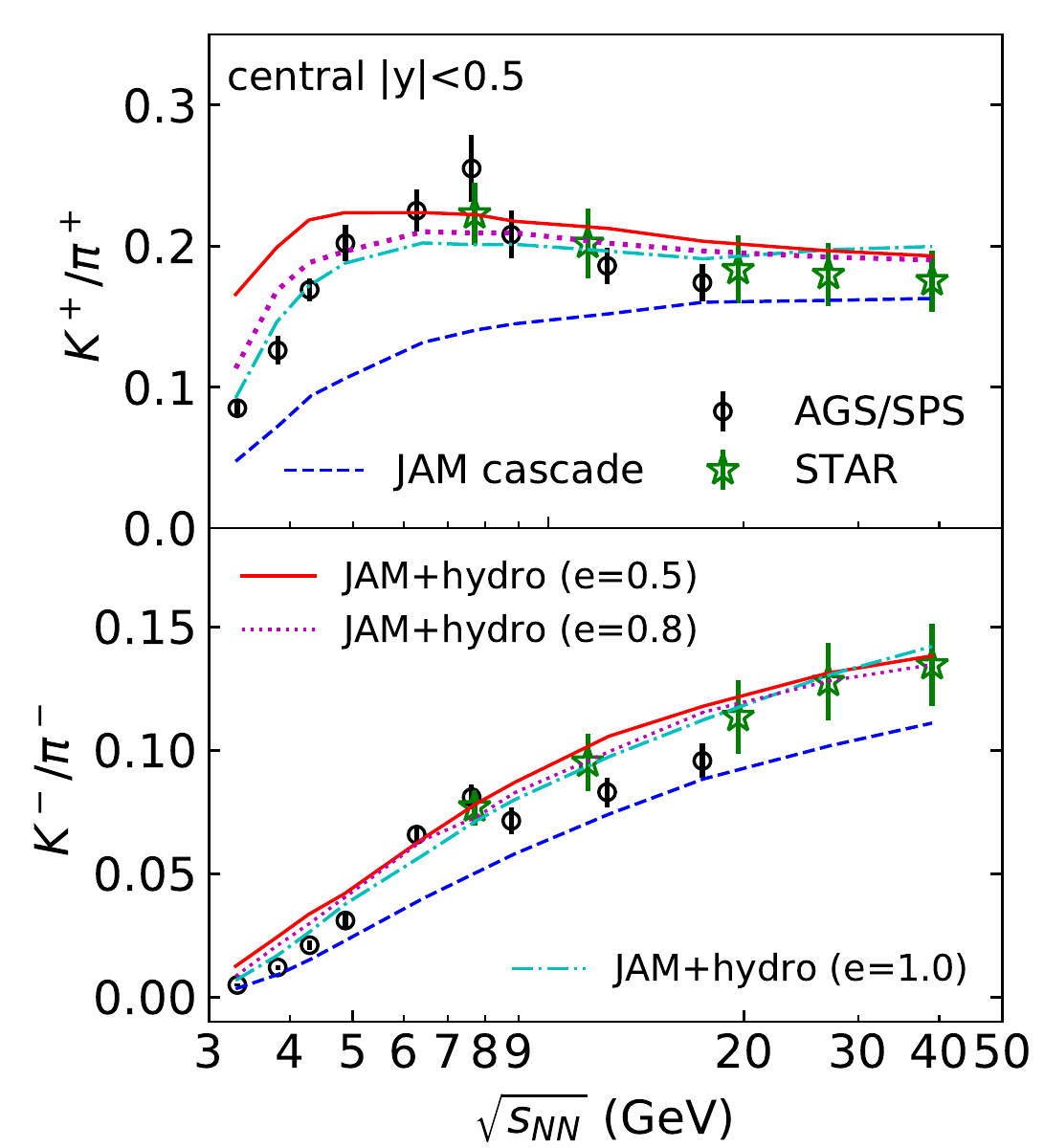}
\caption{Beam-energy dependence of $K/\pi$ ratios at midrapidity
in central Au + Au and Pb + Pb collisions.
The experimental data are taken from Ref.%
~\cite{Adamczyk:2017iwn,Afanasiev:2002mx,Alt:2007aa}.
}
\label{fig:kpi}
\end{figure}
%%%%%%%%%%%%%%%%%%%%%%%%%%%%%%%%%%%%%%%%%%%%%%%%%%%%%%%%%

Let us turn now to the discussion of strange particle to pion ratios.
Beam energy dependence of the $K/\pi$ ratios is shown in Fig.~\ref{fig:kpi}.
Our hybrid approach significantly improves the description of the data
over the predictions from JAM cascade.
To see the dependence on the fluidization energy density
$e_\text{f}$,
we tested three different values of $e_\text{f}=0.5, 0.8$, and 1.0 GeV/fm$^3$.
Larger values of $e_\text{f}$ improve the description of $K/\pi$ ratios
at AGS energies, while it does not affect the higher beam energies much.
This suggests that the value of $e_\text{f}=0.5$ GeV/fm$^3$
may overestimate the thermal part of the system at lower beam energies.
In the 3FD model, the strange particle productions at low beam energies
are overestimated~\cite{Ivanov:2013yla,Batyuk:2016qmb}.  They introduced
a beam energy dependent phenomenological strangeness suppression
factor $\gamma_s$ at $\srtNN \leq5$ GeV to
suppress the yield of strange hadrons.
It was found in the UrQMD hybrid approach
that implementation of the core-corona separation of the system
reduces the ratios involving strange particles~\cite{Steinheimer:2011mp}.
The statistical models can reproduce the structure of the $K^+/\pi^+$
ratio by taking into account canonical suppression
of the strangeness~\cite{Andronic:2005yp,Andronic:2008gu}
or introducing the strangeness suppression factor for the deviation
from chemical equilibrium~\cite{Becattini:2003wp,Becattini:2005xt}.
In our approach, 
good agreement of the $K/\pi$ ratio with the data
is achieved by the partial thermalization in both space and time.
Note that in hybrid approaches,
the nonthermal part of the system, which is described by
the JAM transport model, is in general neither
chemically nor kinetically equilibrated.
Inclusion of the effects of thermal part is essential
for a good description of the $K/\pi$ ratio
within a dynamical model based on hydrodynamics.

We note that the PHSD transport model reproduces
the $K/\pi$ ratio by the strange particle enhancement
due to chiral symmetry restoration
at high baryon densities without assuming any thermalization
\cite{Cassing:2015owa}.
As a future work, incorporation of
the dynamical treatment of a first-order chiral phase transition in a
nonequilibrium real-time dynamics~\cite{Paech:2003fe,Herold:2013qda}
into our framework may improve significantly the description
of collision dynamics and may show some signal of a QGP phase transition
as well as the chiral symmetry restoration for some observables.

%%%%%%%%%%%%%%%%%%%%%%%%%%%%%%%%%%%%%%%%%%%%%%%%%%%%%%%%%
\begin{figure}[tbh]
\includegraphics[width=8.0cm]{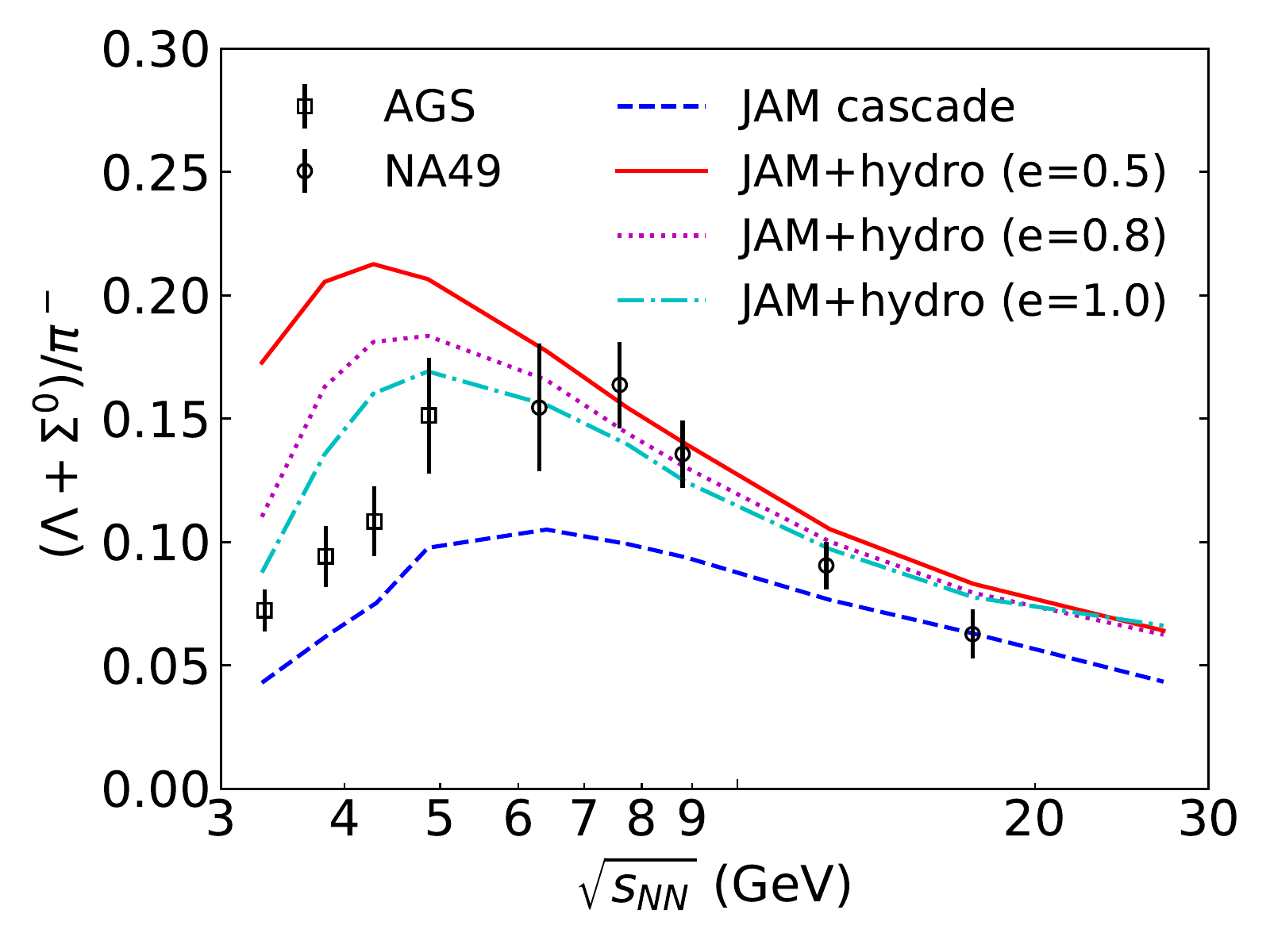}
\caption{Beam energy dependence of $(\Lambda+\Sigma^0)/\pi^-$
ratio at midrapidity
in central Au + Au and Pb + Pb collisions.
The experimental data are taken from Ref.%
~\cite{Anticic:2003ux,Aggarwal:2010ig}.
}
\label{fig:lambdapi}
\end{figure}
%%%%%%%%%%%%%%%%%%%%%%%%%%%%%%%%%%%%%%%%%%%%%%%%%%%%%%%%%

Finally, the beam-energy dependence of $(\Lambda+\Sigma^0)/\pi^-$ ratio
is shown in Fig.~\ref{fig:lambdapi} obtained by
different values of fluidization energy densities.
Good description of the data is obtained at high beam energies
above $\srtNN = 5$ GeV,
but there is an overestimation at lower energies by the hybrid model,
which may suggest that
the conditions of the fluidization depend on beam energy
or additional suppression of the strangeness
due to nonchemical equilibration in the fluid part.

\section{Conclusion}
\label{sec:conclusion}

We have developed a dynamically integrated transport model
for heavy-ion collisions at high baryon densities,
in which nonequilibrium dynamics is solved by the hadronic
transport model,
and a dense part of the system is simultaneously described
by hydrodynamical evolution.
In this approach, dynamical coupling is implemented
through the source terms of the fluid equations. 
For the fluidization of particles, we take into account core-corona
separation, where only the high-density part of the system (core) follows
the hydrodynamical evolution.
We demonstrate that our integrated dynamical approach describes well
the experimental data on the particle yields, transverse mass distributions,
and particle ratios for a wide range of beam energies
$3<\srtNN<30$ GeV
for central Au + Au and Pb + Pb collisions.
We found that partial thermalization of the system is very important
to explain strange particle-to-pion ratios, $K^\pm/\pi^\pm$
and $(\Lambda+\Sigma^0)/\pi^-$.

As future studies,
we plan to perform systematic studies of
the centrality dependence of various observables
including multistrange particles and antibaryons.
The EoS dependence on the particle productions as well as
viscous effects in the hydrodynamical evolution
should be investigated.
In this work, we do not consider possible fluid-particle interactions;
particles that go through dense region are likely to deposit their energies
to the fluid or be absorbed by the fluid. 
These effects may become important for the quantitative description
of some observables.
It is also interesting to look at anisotropic flows such as directed and
elliptic flows within our approach, which are expected to be very
sensitive to the collision dynamics.

\begin{acknowledgments}
This work was supported in part by the
Grants-in-Aid for Scientific Research from JSPS
(Nos.~JP15K05079,  %, %AO & YN (Kiban C)
JP15K05098, %((C TK & YN ) )), 
JP17K05448,
JP17H02900,
JP26220707,
JP17K05438,
and JP17K05442
).
K. Morita acknowledges support by the Polish National Science Center NCN under
Maestro grant DEC-2013/10/A/ST2/00106
and by RIKEN iTHES project and iTHEMS program.
\end{acknowledgments}

\end{document}